%% file: main.tex
\documentclass[conference]{IEEEtran}
\IEEEoverridecommandlockouts
\usepackage{cite}
\usepackage{amsmath,amssymb,amsfonts}
\usepackage{algorithmic}
\usepackage{graphicx}
\usepackage{textcomp}
\usepackage{xcolor}
\usepackage{array}
\usepackage{makecell}
\usepackage{tabularx}
\usepackage{xurl}
\usepackage{caption}
\usepackage{tabularx}
\usepackage{array}
\usepackage{makecell}
\usepackage{caption}
\usepackage{listings}
\usepackage{xcolor}
\usepackage{tcolorbox}
\tcbuselibrary{breakable}

\newcolumntype{L}[1]{>{\raggedright\arraybackslash}m{#1}}
\newcolumntype{Y}[1]{>{\raggedright\arraybackslash}m{#1}}
\newcolumntype{C}[1]{>{\centering\arraybackslash}m{#1}}
\def\BibTeX{{\rm B\kern-.05em{\sc i\kern-.025em b}\kern-.08em
    T\kern-.1667em\lower.7ex\hbox{E}\kern-.125emX}}
\begin{document}

\title{Tailored Prompts, Targeted Protection: Vulnerability-Specific LLM Analysis for Smart Contracts}

\author{
\IEEEauthorblockN{Xing Zhang}
\IEEEauthorblockA{\textit{NetX Foundation}\\
Zug, Switzerland \\
xing.zhang@netx.world}
\and
\IEEEauthorblockN{Keyu Zhang}
\IEEEauthorblockA{\textit{Department of Computer Science}\\
\textit{University of Oxford}\\
Oxford, UK \\
keyu.zhang@cs.ox.ac.uk}
\and
\IEEEauthorblockN{Taohong Zhu}
\IEEEauthorblockA{\textit{Department of Computer Science}\\
\textit{The University of Manchester}\\
Manchester, UK \\
taohong.zhu@manchester.ac.uk}
\and
\IEEEauthorblockN{Anbang Ruan}
\IEEEauthorblockA{\textit{NetX Foundation}\\
Zug, Switzerland \\
ruan@netx.world}
}


\maketitle

\begin{abstract}
Smart contracts on blockchains are prone to diverse security vulnerabilities that can lead to significant financial losses due to their immutable nature. Existing detection approaches often lack flexibility across vulnerability types and rely heavily on manually crafted expert rules. In this paper, we present an LLM-based framework for practical smart contract vulnerability detection. We construct and release a large-scale dataset comprising 31,165 professionally annotated vulnerability instances collected from over 3,200 real-world projects across 15 major blockchain platforms. Our approach leverages precise AST-based context extraction and vulnerability-specific prompt design to instantiate customized detectors for 13 prevalent vulnerability categories. Experimental results demonstrate strong effectiveness, achieving an average positive recall of 0.92 and an average negative recall of 0.85, highlighting the potential of carefully engineered contextual prompting for scalable and high-precision smart contract security analysis.
\end{abstract}


\section{Introduction}
\label{section:intro}
Smart contracts are self-executing programs deployed on blockchain platforms that manage digital assets, facilitate decentralized applications, and execute logic without intermediaries \cite{bartoletti2017empirical}. Their transparency, immutability, and automated execution make them powerful tools. However, these same properties also pose significant security risks, as any flawed logic becomes permanently embedded on-chain and can lead to irreversible financial losses \cite{vacca2021systematic,destefanis2018smart,zhou2023sok}. Moreover, the diversity and complexity of application-specific logic make it extremely difficult to ensure that smart contracts are secure \cite{singh2020blockchain}. As a result, analyzing smart contract vulnerabilities has become a critical area of research, giving rise to a wide range of detection methods and tools.

Traditional approaches to smart contract vulnerability detection, such as formal verification, static analysis, and dynamic analysis, rely heavily on predefined rules. These methods often lack generalizability, are limited to specific vulnerability types, and may require expert knowledge to operate effectively \cite{zhang2023demystifying}..

Recent advances in large language models (LLMs), such as GPT \cite{achiam2023gpt}, have demonstrated strong capabilities in understanding and reasoning about source code \cite{jiang2024survey}. When guided by carefully designed prompting strategies \cite{radford2019language,brown2020language,wei2022chain}, these models can effectively generalize to a wide range of tasks across diverse domains without task-specific fine-tuning.

Recent advances in large language models (LLMs) such as GPT \cite{achiam2023gpt} have demonstrated strong capabilities in understanding and reasoning about source code \cite{jiang2024survey}. When guided by carefully designed prompting strategies \cite{radford2019language,brown2020language,wei2022chain}, these models can effectively generalize to a wide range of tasks across diverse domains without task-specific fine-tuning. Motivated by this progress, recent studies have begun to investigate the use of LLMs for smart contract analysis, particularly for vulnerability detection \cite{destefanis2018smart}. However, existing LLM-based approaches often exhibit a high false positive rate in real-world settings, as models may incorrectly classify benign smart contract logic as vulnerable, thereby limiting their practical applicability.

In this paper, we present a holistic framework for LLM-based smart contract vulnerability detection that bridges the gap between general-purpose LLMs and real-world security requirements. Specifically, we formulate and investigate three core research questions that underpin the effectiveness, robustness, and practicality of LLMs for smart contract vulnerability detection:

\textbf{Q1. What capabilities are required for an LLM-based vulnerability detector?}
To ensure applicability in real-world settings, we first characterize the essential capabilities of an LLM-based detector. These include understanding typical smart contract project structures, contract organization, relevant vulnerability categories, and the appropriate scope of contextual information. This analysis establishes the foundation for the subsequent system design.

\textbf{Q2. How to deliver precise context to the LLM?} By precise, we refer to providing sufficient but not excessive information. The delivered context should supply adequate evidence for accurate vulnerability detection while avoiding redundant or irrelevant details that may distract the model or degrade its reasoning performance.

\textbf{Q3. How can the reasoning ability of a general-purpose LLM be enhanced for vulnerability detection?}
General-purpose LLMs are not explicitly trained for smart contract auditing, which limits their baseline performance in this domain. We therefore explore lightweight enhancement strategies that improve the model’s reasoning capability without requiring complex task-specific pretraining or fine-tuning.

To answer these research questions, we adopt a data-driven and analysis-oriented approach. We first conduct large-scale real-world data collection to construct a comprehensive smart contract vulnerability corpus, which grounds our study in practical security scenarios. Building on this corpus, we design a novel code analysis pipeline for context filtering and construction, enabling the LLM to receive vulnerability-relevant information with minimal noise. Finally, we enhance the reasoning capability of general-purpose LLMs through dedicated few-shot In-Context Learning (ICL) with chain-of-thought (CoT) prompting, without relying on task-specific fine-tuning.

In summary, this paper makes the following key contributions:
\begin{enumerate}
\item \textbf{Comprehensive real-world dataset.}
We collect, curate, and open-source a large-scale real-world smart contract dataset containing 31165 annotated vulnerabilities, sourced from 15 major Ethereum-compatible blockchains, including Ethereum and Binance Smart Chain. This dataset provides a practical and realistic benchmark for evaluating LLM-based smart contract vulnerability detection in real-world settings.

\item \textbf{Novel LLM-based detector framework.}
We propose an end-to-end vulnerability detection framework that leverages, for the first time, AST-based code analysis to construct structured and vulnerability-aware representations of smart contracts as inputs to LLMs. This design captures both syntactic structure and vulnerability-relevant semantic information, enabling LLMs to reason more effectively about contract logic and architectural relationships. Moreover, we enhance the constructed context through few-shot ICL with CoT prompting, improving the reasoning capability of general-purpose LLMs.

\item \textbf{Vulnerability-aware prompt design.}
Based on extensive empirical analysis and experiments on real-world smart contracts, we design customizable prompting strategies tailored to 13 vulnerability types. Our approach achieves strong detection performance, attaining up to 0.90 recall on real-world test contracts.
\end{enumerate}

\section{Related Work}

\subsection{Traditional Techniques for Smart Contract Vulnerability Detection}
Smart contracts are self-executing programs deployed on blockchains, typically written in high-level languages such as Solidity \cite{wang2018overview}. A variety of traditional techniques have been proposed to analyze Solidity smart contracts and detect security vulnerabilities from different perspectives.
Existing approaches can be broadly classified into formal verification, symbolic execution, fuzzing, intermediate representation–based analysis, and deep learning–based methods \cite{qian2022smart}. 

Formal verification techniques model smart contracts using formal semantics and mathematical logic to verify whether contract executions satisfy predefined security or correctness properties under all possible states, as demonstrated by frameworks such as F* \cite{grishchenko2018semantic} and KEVM \cite{hildenbrandt2018kevm}. In contrast, symbolic execution analyzes smart contract behaviors by treating inputs and state variables symbolically and systematically enumerating feasible execution paths under path constraints; representative tools such as Oyente \cite{luu2016making} and Mythril \cite{mueller2017framework} adopt this technique to detect semantic vulnerabilities. Complementing static reasoning, fuzzing-based methods, including ContractFuzzer \cite{jiang2018contractfuzzer} and Reguard \cite{liu2018reguard}, dynamically generate large numbers of test inputs or transaction sequences and execute smart contracts to expose abnormal or vulnerable runtime behaviors. To improve analyzability, intermediate representation–based approaches, such as Slither \cite{feist2019slither} and Vandal \cite{brent2018vandal}, translate smart contracts into structured representations that preserve control-flow, data-flow, or dependency information, enabling more systematic static analysis. More recently, deep learning–based methods model smart contracts as sequences or graphs and automatically learn vulnerability patterns from data, as exemplified by SaferSC \cite{tann2018towards} and DR-GCN \cite{zhuang2021smart}.

Despite their progress, traditional techniques generally depend on fixed analysis strategies or limited semantic modeling, making it difficult to adapt across diverse vulnerability types or capture high-level program intent in complex smart contracts \cite{zhang2023demystifying,sun2024gptscan}.

\subsection{LLM for Smart Contract Vulnerability Detection}
Large Language Models (LLMs) are neural models with a large number of parameters trained on large-scale textual data, enabling strong capabilities in understanding and generating natural language \cite{chen2021evaluating,kasneci2023chatgpt,zhao2023survey}. Through carefully designed prompts, LLMs can be guided to perform code-related tasks such as program analysis and vulnerability detection without modifying model parameters, a practice known as prompt engineering \cite{sahoo2024systematic}. Zero-shot prompting relies solely on task descriptions \cite{radford2019language}, while few-shot prompting further incorporates a small number of examples to improve performance on complex tasks \cite{brown2020language}. In contrast, Chain-of-Thought prompting explicitly encourages LLMs to decompose a task into intermediate reasoning steps, allowing the model to conduct structured, step-by-step analysis before producing final conclusions \cite{wei2022chain}.

Recent work has explored the use of LLMs for smart contract vulnerability detection, mainly by leveraging their ability to understand code semantics and reason over program logic. Many approaches combine LLMs with external context or program analysis techniques to improve detection quality. Ding et al. \cite{ding2025smartguard} proposed a framework that retrieves semantically similar smart contracts from a historical corpus and includes them in the prompt together with Chain-of-Thought reasoning, allowing the LLM to analyze vulnerabilities based on relevant prior examples. Extending this idea, Zaazaa et al. \cite{zaazaa2024smartllmsentry} used LLMs to generate vulnerability detection rules dynamically, which are then incorporated into static analyzers using structural information such as abstract syntax trees and control-flow graphs, enabling the system to better adapt to newly emerging vulnerability patterns.

Other studies explore different ways of applying LLMs to vulnerability detection. Sun et al. \cite{sun2024gptscan} presented a hybrid approach in which LLMs first identify suspicious code regions and reason about potential logic vulnerabilities, and the results are then verified using static analysis to reduce false positives, especially for business logic issues. In contrast, Boi et al. \cite{boi2024vulnhunt} examined the direct use of ChatGPT with fixed prompts, showing that prompt-only analysis can detect common vulnerabilities but is limited by insufficient context and unstable reasoning. Beyond prompt design, Yang et al. \cite{yang2023automated} fine-tuned pretrained LLMs on vulnerability-labeled audit data to improve detection performance, although the results remain constrained by data scale and vulnerability coverage. More recently, Bouafif et al. \cite{bouafif2024context} proposed a context-driven co-auditing approach that enriches LLM inputs with function-level call relationships and vulnerability-specific instructions, which helps reduce ambiguity and improve the consistency of analysis.

\section{Approach}
\subsection{Overview}
Providing large language models with entire smart contract codebases, together with ill-defined detection objectives, noisy or excessive contextual information, weak prompting strategies, and unstructured outputs, often leads to poor vulnerability detection performance. To make general-purpose LLMs effective for real-world smart contract vulnerability detection, we structure our approach around the three core research questions introduced in Section~\ref{section:intro}.

\begin{figure}[h]
    \centering
    \includegraphics[width=0.9\linewidth]{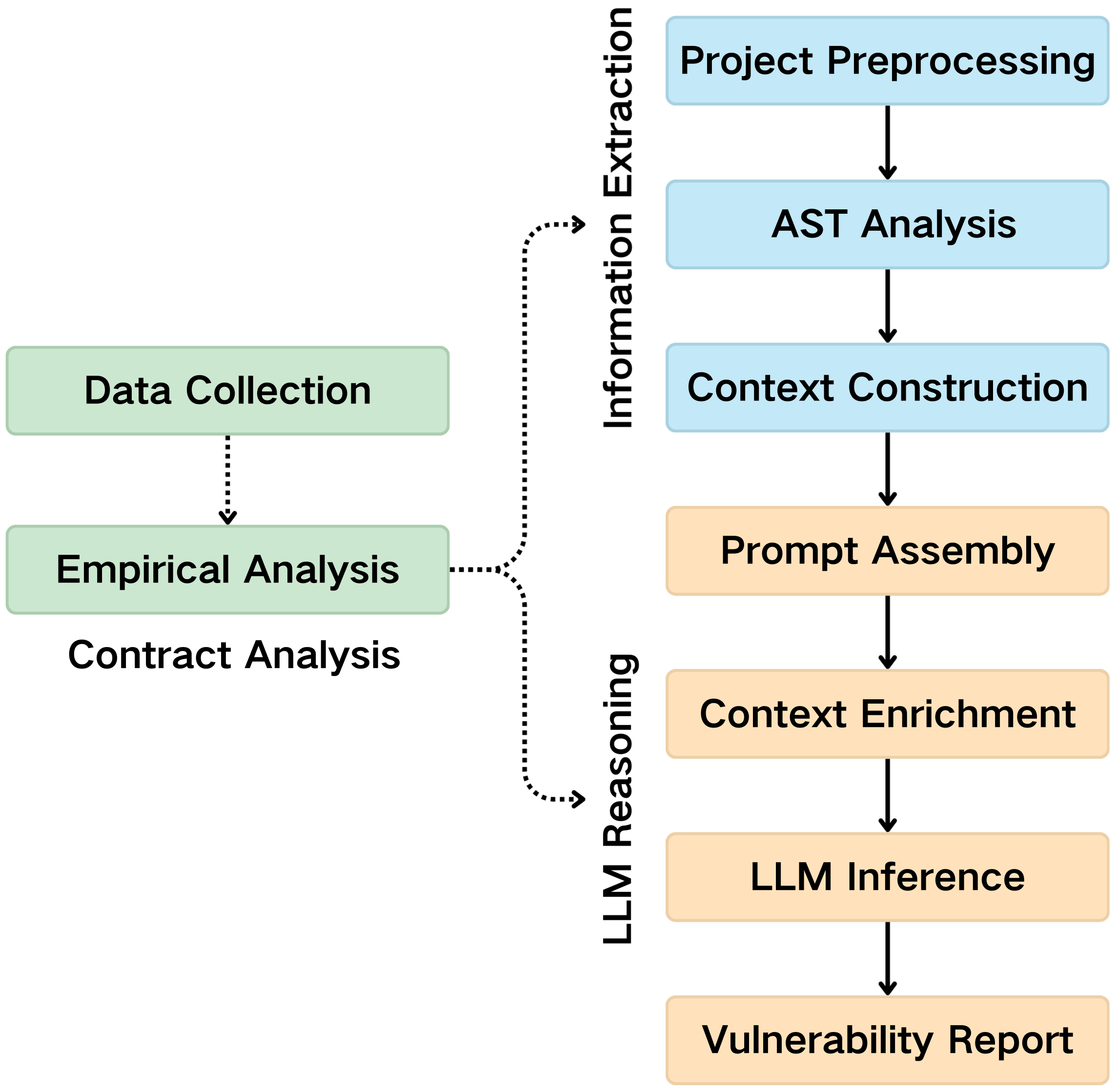}
    \caption{Overview of the proposed LLM-based smart contract vulnerability detection framework.}
    \label{fig:framework}
\end{figure}

Specifically, our framework is illustrated in Fig.~\ref{fig:framework}. We first conduct large-scale real-world data collection from major smart contract auditing platforms across multiple EVM-compatible chains (Data Collection), followed by systematic empirical analysis to characterize vulnerability patterns and auditing requirements (Empirical Analysis).

Building on this corpus, we design a vulnerability detector tailored to realistic contract auditing scenarios. As shown in Fig.~\ref{fig:framework}, the detector workflow begins with project preprocessing, which filters project-level smart contracts to reduce the input scope to security-relevant components. On this pruned contract set, we perform AST-based code analysis to identify structural relationships and extract vulnerability-critical information, enabling precise context construction.

The resulting structured context is then incorporated into a carefully engineered prompt assembly stage and further enhanced through few-shot ICT with CoT prompting (context enrichment), strengthening the LLM’s reasoning capabilities. Finally, the constructed prompts are submitted to the LLM for inference (LLM inference), producing structured and interpretable vulnerability reports.

The following subsections describe each component of our framework in detail.

\subsection{Real-World Data Collection and Analysis}

\begin{table*}[t]
\centering
\renewcommand{\arraystretch}{1.3}
\captionsetup{skip=6pt}
\begin{tabularx}{\textwidth}{L{0.25\textwidth}|Y{0.55\textwidth}|C{0.12\textwidth}}
\hline
\thead{Vulnerability Category} & \thead{Brief Description} & \thead{Number of Cases} \\
\hline
Reentrancy & 
A vulnerability where external calls are performed before internal state updates, enabling attackers to repeatedly re-enter a function and manipulate contract state to drain assets. & 375 \\
\hline
Missing Event & 
The absence of expected event emissions following critical state changes, reducing transaction transparency and hindering off-chain monitoring and auditing. & 1117 \\
\hline
Centralization & Excessive control concentrated in privileged accounts or roles, undermining decentralization assumptions and introducing single points of failure. & 4341 \\
\hline
Input Validation & Insufficient validation or sanitization of external inputs, potentially leading to unintended execution paths or security vulnerabilities. & 822 \\
\hline
Weak Randomness & Reliance on predictable on-chain data sources (e.g., block timestamps or hashes) for randomness generation, allowing adversaries to influence outcomes. & 16 \\
\hline
Sandwich Attack & An economic attack in which adversaries front-run and back-run user transactions to exploit price slippage in decentralized exchanges. & 185 \\
\hline
Redundant Statements & Superfluous or unreachable code segments that increase gas consumption and reduce code clarity without affecting functionality. & 276 \\
\hline
Flashloan Attack & Exploitation of protocol logic using uncollateralized flash loans to manipulate state or pricing within a single transaction. & 24 \\
\hline
Too Many Digits & The use of long or poorly formatted numeric literals that reduce code readability and increase the likelihood of human auditing errors. & 80 \\
\hline
Error Message & Unclear, missing, or misleading error messages that complicate debugging, auditing, and vulnerability diagnosis. & 634 \\
\hline
Constant Optimization & 
Improper use or omission of constant variables, leading to reduced code clarity and suboptimal gas efficiency. & 136 \\
\hline
Return Value Check & Failure to verify function return values, potentially resulting in silent failures or unintended execution states. & 283 \\
\hline
Division before Multiplication & Performing division operations before multiplication in integer arithmetic, causing precision loss due to truncation. & 61 \\
\hline
\end{tabularx}
\caption{Vulnerability categories considered in this study and their distributions in real-world audits.}
\label{tab:list-of-vul}
\end{table*}

To construct a practical LLM-based smart contract auditing tool suitable for real-world settings, it is essential to curate a dataset that pairs real-world source code with professionally produced audit reports. In the Web3 ecosystem, development teams commonly engage third-party security auditors to assess smart contracts and identify potential vulnerabilities. These audit reports are typically released publicly after deployment, providing authoritative documentation of security issues within specific codebases.

Accordingly, we conduct large-scale real-world data collection by crawling audit reports from prominent cryptocurrency gateways and security organizations, including CoinMarketCap, CertiK, QuantStamp, and ConsenSys. As a result, the collected reports span from their inception through September 2023 and cover 15 blockchain networks, such as Ethereum, BNB Chain, and Solana. 

To systematically process these reports, we employ automated PDF analysis techniques to extract key metadata from each audit report, including audited commit hashes, GitHub repository links, and documented vulnerability descriptions. Leveraging the audited commit information, we then retrieve the corresponding source code from GitHub, enabling the construction of a synchronized dataset in which each codebase is explicitly aligned with its associated vulnerability reports.

Through this process, the resulting dataset provides comprehensive coverage of the smart contract security landscape, comprising 2874 Web3 projects, more than 73,000 Solidity (\texttt{.sol}) files, and 31165 documented vulnerability instances across 15 major blockchain platforms. To support reproducibility and facilitate future research, we release this large-scale real-world dataset as open source and make it publicly available at \cite{LLM_VUL_DETECTOR}.

This comprehensive dataset serves as the foundation for the design of our LLM-based vulnerability detector. Through systematic empirical analysis, we examine key characteristics of real-world smart contracts, including project organization, inter-contract relationships, vulnerability types, and their intrinsic properties. These insights not only guide the subsequent system design, but also inform the experimental evaluation of our approach.

In this work, we focus on commonly occurring vulnerabilities observed in real-world audits to ensure the practicality and real-world applicability of the proposed detector. These vulnerability categories are selected based on their prevalence in auditing workflows and the availability of a statistically significant number of samples in our dataset. The selected categories, together with their formal definitions and instance counts, are summarized in Table~\ref{tab:list-of-vul}.


\subsection{AST Information Extraction}
Directly providing raw smart contract source code to LLMs for vulnerability detection presents significant challenges. As probabilistic models, LLMs often struggle to identify subtle vulnerability-related patterns when they are embedded within complex surrounding logic. However, accurate vulnerability detection requires a high degree of precision, where even minor code modifications may determine the presence or absence of a security flaw.

To address this limitation, we adopt symbolic and structural representations of source code to reduce redundancy and expose vulnerability-relevant semantics. Specifically, we parse smart contract source code into an AST and leverage vulnerability-specific contextual knowledge to extract only the code fragments most relevant to potential flaw locations. This design enables the construction of precise and focused inputs for subsequent LLM-based analysis.

\begin{figure}[t]
    \centering
    \includegraphics[width=1\linewidth]{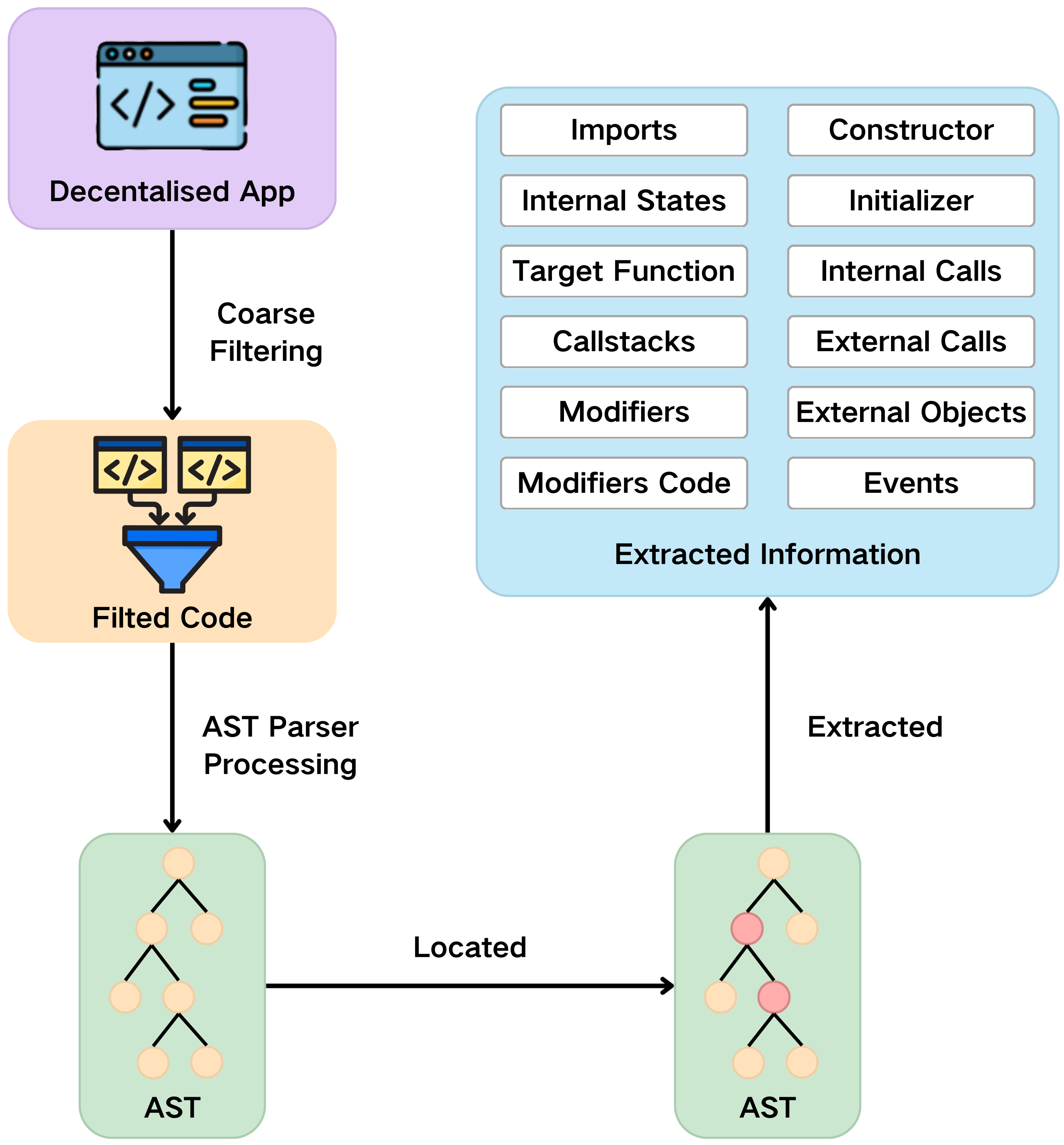}
    \caption{AST information extraction overview}
    \label{fig:AST_code}
\end{figure}

Based on this representation, we introduce a two-stage code processing pipeline consisting of coarse filtering and fine-grained extraction to mitigate both context overload and under-specification, while preserving the analytical fidelity of the LLM, as illustrated in Figure~\ref{fig:AST_code}. The extracted information is then formatted and organized for downstream prompt construction.

Given a smart contract project and a target function associated with a specific vulnerability detection task, the coarse filtering stage leverages the project’s organizational structure to retain only core smart contracts. This step restricts the initial analysis scope by excluding auxiliary components that are unlikely to contain vulnerability-relevant logic.

Building on extensive empirical analysis of the collected dataset, we observe that most common smart contract vulnerabilities are localized within specific functions, depending on the vulnerability type. For example,  Reentrancy vulnerabilities typically occur in functions that invoke external contracts, whereas timestamp-dependence vulnerabilities arise in functions that rely on \texttt{block.timestamp} for critical computations or control-flow decisions. Guided by these observations and expert knowledge from real-world auditing practice, the fine-grained extraction stage extracts call-stack functions and associated contextual information most relevant to the targeted vulnerability, further narrowing the analysis scope.

To support fine-grained extraction, the retained smart contract project is comprehensively analyzed using an AST-based pipeline, as illustrated in Figure~\ref{fig:AST_code}. Specifically, we employs python-solidity-parser \cite{solidity_parser} to parse the entire smart contract project to construct a symbolic abstract syntax tree (AST), instantiated as multiple interlinked syntax trees in memory. Based on this representation, the parser locates the target function and identifies the syntax tree corresponding to its enclosing contract. Starting from this location, vulnerability-relevant contextual information is extracted, with code snippets originating from candidate functions. This preprocessing step ensures that each vulnerability instance is represented by a structured, AST-derived code fragment suitable for downstream analysis, covering the following categories:


\begin{itemize}
    \item \textbf{Imports:} External files or libraries imported and referenced by the target function.
    \item \textbf{Internal States:} All state variables accessed or modified by the target function.
    \item \textbf{Target function:} The complete source code of the function.
    \item \textbf{Callstack:} The function call relationships originating from the target function, including parameters and return values.
    \item \textbf{Modifiers:} Modifiers applied to the target function.
    \item \textbf{Modifiers codes:} The full source code of the applied modifiers.
    \item \textbf{Constructor:} The constructor code of the enclosing smart contract.
    \item \textbf{Initializer:} The initializer logic of the smart contract, if present.
    \item \textbf{Internal calls:} Internal functions invoked by the target function.
    \item \textbf{External calls:} External contract functions invoked by the target function.
    \item \textbf{External object:} External contracts or addresses interacted with by the target function.
    \item \textbf{Event:} Event definitions and emissions associated with the target function.
\end{itemize}

After completing the two-stage processing pipeline, the extracted information is organized into a flattened JSON-formatted representation suitable for LLM consumption, thereby completing the construction of precise context for the target function within the specific vulnerability-detection task.

\subsection{Context-Enriched LLM Detection}
Based on our empirical analysis and experimental results, we design a prompt template that provides a well-defined and information-rich context for smart contract vulnerability detection. We adopt this unified prompt template applicable across multiple vulnerability categories, which is dynamically instantiated with vulnerability-specific information at inference time. The template consists of the following components:

\begin{figure}[b]
    \centering
    \includegraphics[width=0.6\linewidth]{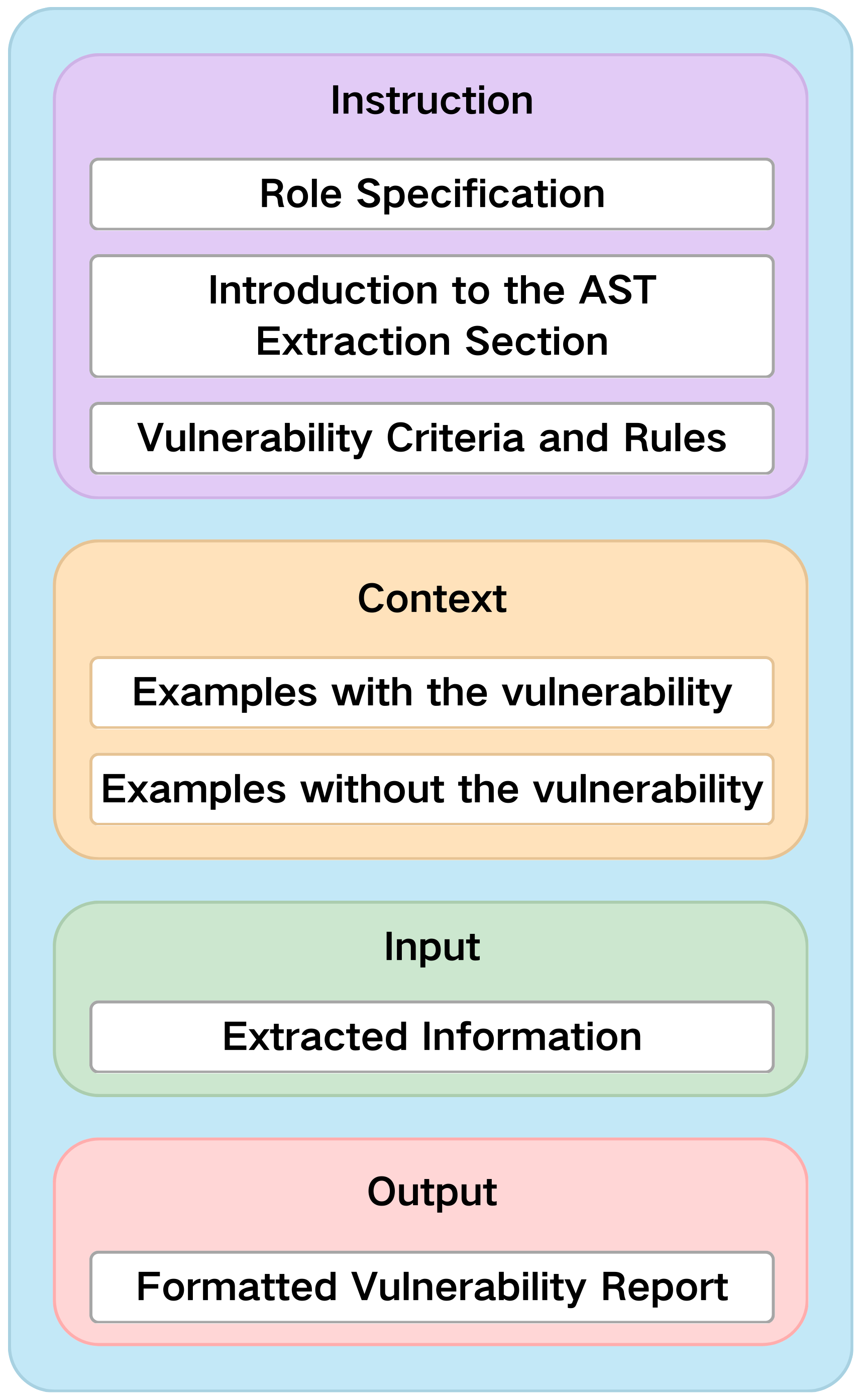}
    \caption{Prompt template design}
    \label{fig:prompt}
\end{figure}

\input{TableExp}

\begin{itemize}
    \item \textbf{Instruction}: This component defines the LLM’s role as a smart contract auditor and specifies the expected input structure, as described in the AST information extraction stage. It also outlines common vulnerability indicators and heuristic rules to guide the detection process. The detailed instruction component of the prompt template, instantiated for reentrancy vulnerability detection as a representative example, is provided in Appendix~\ref{app:prompt}.
    \item \textbf{Context}: 
    To enable few-shot ICL, this section includes representative positive and negative examples of the target vulnerability. These examples are manually curated by security experts, and each vulnerability category is associated with a distinct set of examples that are selected based on the detection task.
    \item \textbf{Input}: This component contains the AST-parsed and filtered code fragments extracted from the target contract. The LLM analyzes these snippets to determine the presence of vulnerabilities and identify the corresponding vulnerable code locations.
    \item \textbf{Output}: This section specifies the required response format, including the vulnerability verdict, the line number of the detected issue, and the relevant code fragment.
\end{itemize}

During inference, the detection system instantiates the prompt by populating the Context and Input components with vulnerability-specific examples and the corresponding code segment, before submitting the finalized prompt to the LLM for prediction.

As the input construction process has been described in the previous subsection, we focus here on the design of the Context component, which plays a critical role in enriching the prompt with accurate vulnerability definitions and representative reference cases.

In this context, we select contextual examples for the LLM based on both empirical analysis of the collected dataset and experimental observations. For each vulnerability category, we first construct an initial set of representative examples that clearly exhibit the target vulnerability and evaluate the detection performance using these examples. If the performance falls below the target benchmark, we conduct a manual review of misclassified cases to refine the prompt templates. Through this refinement process, we identify boundary cases that are prone to misclassification by the LLM, which typically involve subtle or ambiguous patterns frequently confused with benign logic. These boundary cases are subsequently incorporated into the contextual examples by adding new reference instances and enriching the associated explanations with more explicit, step-by-step reasoning. This evaluation–refinement process is iteratively repeated until no further significant performance improvements are observed, resulting in a balanced set of contextual examples.

Throughout this iterative process, the contextual examples are structured to pair source code with detailed explanations of the underlying vulnerability logic and corresponding decision criteria. These explanations provide implicit CoT guidance by exposing the intermediate reasoning steps used to assess vulnerability presence. In this design, true-positive examples guide the model toward correct vulnerability identification, whereas carefully selected false-negative examples help reduce over-detection. This balanced and reasoning-enriched prompt construction is essential for achieving reliable performance in real-world deployments and maintaining a favorable user experience in practical smart contract auditing scenarios.

To evaluate the effectiveness of incorporating detailed explanation–based CoT reasoning in our design, we conduct a comparative experiment that contrasts prompt configurations with and without CoT reasoning. Our results show that incorporating CoT reasoning improves detection performance over standard few-shot prompting, imporve from 0.85 to 0.91 for all vunerabilities in averay, achieving the accuracy levels required for the system’s objectives.

To evaluate the effectiveness of incorporating explanation-based CoT reasoning, we conduct a comparative experiment
that contrasts prompt configurations with and without CoT reasoning. Our results show that incorporating CoT reasoning improves detection performance over standard few-shot prompting, increasing the average accuracy across vulnerability types from 0.85 to 0.91, which contributes to improved overall detection reliability.


\section{Experiment}
For each vulnerability category, we randomly construct an evaluation dataset by sampling both positive and negative instances from the entire collected dataset. Positive instances correspond to entries labeled with the target vulnerability, whereas negative instances are entries that do not exhibit this vulnerability. To ensure sufficient coverage, the number of negative instances is set to be equal to or up to twice the number of positive instances.

Due to the limited availability of certain vulnerability types, positive instances may partially overlap with those used during prompt design. In contrast, negative instances are sampled from a substantially larger pool and are therefore varied extensively, leveraging the size and diversity of the dataset to reduce sampling bias and ensure the reliability of the evaluation results.

Following data selection, each instance is independently processed by our automatically developed analysis pipeline and embedded into the vulnerability-specific prompt template. The resulting prompts are submitted to the OpenAI API using the GPT-4 model for inference. To comply with API constraints, the complete prompt, including role instructions, contextual examples and AST-derived source code snippets, is restricted to fewer than 4,000 tokens.

Lastly, the model outputs are aggregated to compute recall metrics for both positive and negative instances, referred to as positive recall and negative recall, respectively. Both metrics are equally important in practice. In particular, a high negative-recall rate corresponds to a low false-positive rate, which is critical for the practical deployment of vulnerability detectors in real-world auditing workflows, as each false-positive detection typically requires manual verification by security auditors, incurring substantial time and resource costs. This practical constraint has not been adequately addressed by prior LLM-based approaches. To facilitate comparison with previous work, we additionally give precision and accuracy metrics for completeness.

We report validation results for all 13 vulnerability categories considered in this study, with detailed results presented in Table~\ref{tab:list-detection-rate}.







\section{Conclusion}
This paper presents a practical and effective framework for adapting general-purpose LLMs to smart contract vulnerability detection. We identify three fundamental research questions and address them through a comprehensive study. Specifically, we define the required capabilities of an LLM-based detector based on large-scale real-world data analysis, construct precise and vulnerability-aware contexts using AST-based code analysis, and enhance model reasoning through tailored few-shot in-context learning with chain-of-thought prompting. Through these efforts, we bridge the gap between powerful general-purpose LLMs and the specialized requirements of smart contract security.

Our contributions include an open-sourced large-scale dataset comprising 31165 professionally annotated vulnerabilities from real-world audits across 15 major blockchain platforms, a novel end-to-end vulnerability detection pipeline, and customized prompt templates for 13 prevalent vulnerability categories. Experimental results demonstrate strong performance, achieving positive recall rates ranging from 0.76 to 1.00 with an average of 0.92, and negative recall rates ranging from 0.60 to 1.00 with an average of 0.85. These results validate the effectiveness and practical applicability of the proposed approach in real-world settings.

While the reliance on manual prompt refinement limits the ability to generalize to previously unseen vulnerability patterns, this work lays a solid foundation for future improvements, such as automated and adaptive prompt optimization techniques. Overall, our framework advances accessible and high-precision tools for securing decentralized applications and protecting digital assets in blockchain ecosystems.

\input{appendix}

\bibliographystyle{ieeetr}
\bibliography{all}

\end{document}

%% file: TableExp.tex
\begin{table*}[!tb]
\centering
\renewcommand{\arraystretch}{1.3}
\captionsetup{skip=6pt}
\begin{tabularx}{\textwidth}{l|
>{\centering\arraybackslash}X|
>{\centering\arraybackslash}X|
>{\centering\arraybackslash}X|
>{\centering\arraybackslash}X}
\hline
\thead{Vulnerability Type} &
\thead{Positive Recall} &
\thead{Negative Recall} &
\thead{Precision} &
\thead{Accuracy} \\
\hline
Reentrancy                     & 0.76 & 0.85 & 0.91 & 0.79 \\
Missing Event                  & 0.98 & 0.60 & 0.84 & 0.86 \\
Centralization                 & 0.95 & 0.80 & 0.94 & 0.92 \\
Input Validation               & 0.93 & 0.68 & 0.86 & 0.85 \\
Weak Randomness                & 0.73 & 1.00 & 1.00 & 0.91 \\
Sandwich Attack                & 0.89 & 1.00 & 1.00 & 0.94 \\
Redundant Statements           & 0.92 & 0.90 & 0.84 & 0.91 \\
Flashloan Attack               & 1.00 & 0.92 & 0.88 & 0.95 \\
Too Many Digits                & 1.00 & 0.78 & 0.83 & 0.89 \\
Error Message                  & 0.90 & 0.98 & 0.98 & 0.94 \\
Constant Optimization          & 0.92 & 0.96 & 0.90 & 0.95 \\
Return Value Check             & 0.99 & 0.96 & 0.97 & 0.98 \\
Division before Multiplication & 0.98 & 0.86 & 0.88 & 0.92 \\
\hline
\textbf{Average}               & \textbf{0.92} & \textbf{0.87} & \textbf{0.91} & \textbf{0.91} \\
\hline
\end{tabularx}
\caption{Detection performance across different vulnerability categories evaluated on real-world smart contract audits.}
\label{tab:list-detection-rate}
\end{table*}

%% file: appendix.tex
\appendices
\section{Instruction Component of the Prompt Template}
\label{app:prompt}

This appendix provides the instruction component of the prompt template
used for reentrancy vulnerability detection as a representative example.
For other vulnerability types, the same prompt structure is employed,
with only the vulnerability-specific information modified
accordingly.

\begin{tcolorbox}[breakable, 
title={Instruction Prompt for Reentrancy Detection}]
Act as a Smart Contract Security auditor. Your task is to analyze a Solidity smart contract inside the \textless code\textgreater \space tag to identify whether it contains a direct or indirect risk of a reentrancy attack. You must treat text inside ** with more importance and attention.

The smart contract is provided as follows:

- Imports: this section will contain import statements in the file

- Internal states: this section will contain the internal states of the smart contract

- Target function: this section will contain the target function of the smart contract where the reentrancy attack might originate from or happen

- Callstack: this section will contain the callstack of the target function. The callstack functions might contain reentrancy attacks.

- Modifiers: this section will contain the modifiers of the smart contract

- Modifiers codes: this section will contain the modifiers codes of the smart contract

- Constructor: this section will contain the constructor code of the smart contract

- Initializer: this section will contain the initializer code of the smart contract

- Internal calls: this section will contain the internal function calls of the target function

- External calls: this section will contain the external function calls of the target function

- External objects: this section will contain the external objects of the smart contract

- Events: this section will contain the events emitted by the smart contract

For each reentrancy attack, provide a json object with the following keys enclosed in double quotes: is\_vulnerable, and code\_snippet. If an attack is not present, provide a json object with the following key: is\_vulnerable.

Any external function call may lead to a reentrancy attack if invoked before a state change or events emission.
A function called with super is not an external call as it is executed in the same context as the caller.

**A target function using a nonReentrant modifier is not vulnerable to reentrancy attacks.**

Common indicators of reentrancy attacks include:

- Use of low-level calls such as call or sendValue

- Internal state changes after external calls

- Lack of use of the checks-effects-interactions pattern

- Lack of use of mutexes or reentrancy guards(eg. nonReentrant modifier)

- Emitting events after external calls

A few important rules:

- Events are not considered as external calls.

- If an external call is the last statement in the target function, then it is not vulnerable to reentrancy attacks.

- transfer() and send() are safe given that they limit the gas forwarded to the recipient to 2300 gas, which is not enough to call back into the calling contract.

- If there are any internal state changes after external calls, then set is\_vulnerable to true. If there is no such vulnerability, then set is\_vulnerable to false.
\end{tcolorbox}